\documentclass[twocolumn,showpacs,showkeys,preprintnumbers,superscriptaddress]{revtex4-1}
\usepackage{amssymb}

\usepackage{amsfonts}

\usepackage{latexsym}

\usepackage{dcolumn}

\usepackage{amsmath}

\usepackage{graphicx}

\newcommand{\eqn}{\ref}

\begin{document}

\title{Entangling optical and microwave cavity modes by means of
a nanomechanical resonator}
\author{Sh. Barzanjeh}
\affiliation{Department of Physics, Faculty of Science,
University of Isfahan, Hezar Jerib, 81746-73441, Isfahan, Iran}
\affiliation{School of Science and Technology, Physics Division, Universit\`{a} di Camerino,
I-62032 Camerino (MC), Italy}
\author{D. Vitali}
\affiliation{School of Science and Technology, Physics Division, Universit\`{a} di Camerino,
I-62032 Camerino (MC), Italy}
\author{P. Tombesi}
\affiliation{School of Science and Technology, Physics Division, Universit\`{a} di Camerino,
I-62032 Camerino (MC), Italy}
\author{G. J. Milburn}
\affiliation{Centre for Engineered Quantum Systems, School of Physical Sciences, The
University of Queensland, Saint Lucia, QLD 4072, Australia}

\date{\today}

\begin{abstract}

We propose a scheme able to generate stationary continuous variable entanglement between an optical and a microwave cavity mode by means of their common interaction with a micro-mechanical resonator.  We show that when both cavities are intensely driven one can generate bipartite entanglement between any pair of the tripartite system, and that, due to entanglement sharing, optical-microwave entanglement is efficiently generated at the expense of microwave-mechanical and opto-mechanical entanglement.
\end{abstract}


\maketitle

\section{Introduction}\label{Introduction}

Micro- and nano-mechanical resonators can be efficiently coupled with a large number of different devices and therefore represent a key tool for the realization of quantum interfaces, able to store and redistribute quantum information. Very popular examples are electromechanical systems in which a mechanical resonator (MR) can be coupled either capacitively or inductively with a an electric circuit~\cite{blencowe,roukpt,tianzoller}, or optomechanical systems in which either radiation pressure or the gradient force allows strong coupling with an optical cavity mode \cite{kippenberg,amo,marquardt0}. More recently various schemes have been proposed in order to couple a MR either with single atoms \cite{atom-membrane} or with atomic ensembles \cite{Reichel07,fam,ian-ham,treutlein}. A quantum interface must be able to transfer with high fidelity quantum state between different degrees of freedom and this can be implemented either via a quantum transfer protocol or exploiting entanglement. A large number of schemes have been recently proposed for entangling hybrid systems involving a MR. One could entangle a nanomechanical oscillator with a Cooper-pair box \cite{Armour03}, or may entangle two charge qubits \cite{zou1} or two Josephson junctions \cite{cleland1} via nanomechanical Alternatively, schemes for entangling a superconducting coplanar waveguide field with a nanomechanical resonator, either via a Cooper pair box within the waveguide \cite{ringsmuth}, or via direct capacitive coupling \cite{Vitali07}, have been proposed.

Motivated by the above-mentioned studies, we show that a MR can be employed to couple efficiently microwave and optical fields, which simultaneously interact with it. In particular we show that stationary, i.e., long-lived, microwave-optical entanglement can be generated in such a hybrid tripartite device. This mechanical transduction at the quantum level would be extremely useful in quantum information networks because light modes are unaffected by thermal noise and easily connect distant nodes of the network, while microwave cavities in each nodes can be efficiently coupled with molecular and solid-state qubits \cite{Rabl2006,Blais2007}.

The considered hybrid system was also recently suggested in Ref.~\cite{Regal2011} and could be based on a lumped-element superconducting circuit with a free standing drum-head capacitor analogous to that of Ref.~\cite{Teufel}. In fact, the drum-head capacitor could be optically coated and form one micro-mirror of a Fabry-Perot optical
cavity. In this way the vibrating drum-head is at the same time capacitively coupled with the microwave cavity mode and coupled via radiation pressure with the optical cavity mode.

The paper is organized as follows. In Sec. II we describe the proposed system and derive the quantum
Langevin equations (QLE). In Sec. III, the linearization of the quantum Langevin equations around the semiclassical steady
state is discussed. In Sec. IV we study the steady state of the system and quantify the entanglement by using the logarithmic negativity. Our conclusions are summarised in Sec. V.

\begin{figure}[ht]

\includegraphics[width=0.45\textwidth]{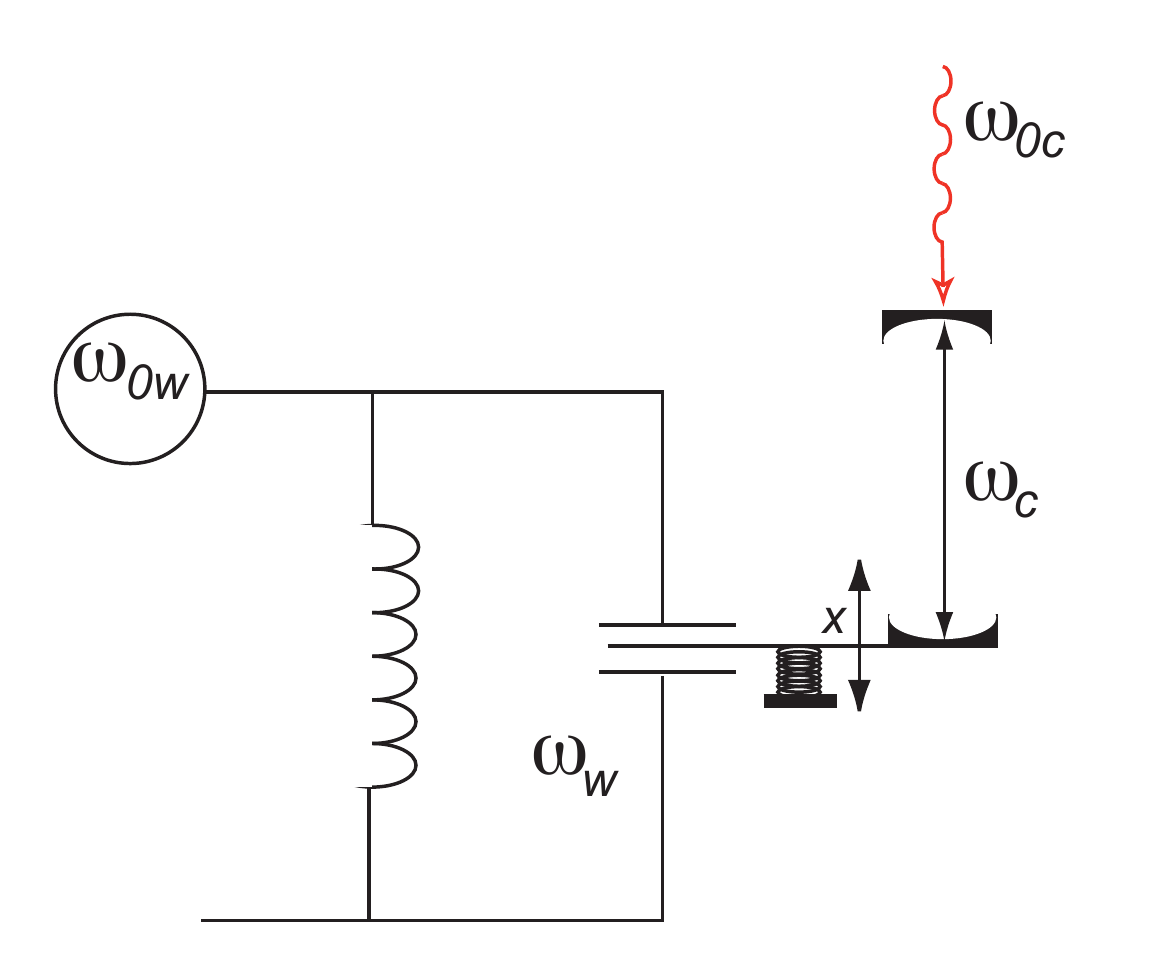}
\caption{Schematic description of the device under study: a lumped-element microwave cavity with a free standing drum-head capacitor is coupled also to an optical cavity formed by an input mirror and the optically coated drum-head capacitor. }
\label{fig:1}
\end{figure}

\section{System dynamics}

The proposed scheme is shown in Fig.~1. We assume a MR which on the one side is capacitively coupled to the field of
a superconducting microwave cavity (MC) of resonant frequency $\omega_w$ and, on the other side,
coupled to a driven optical cavity (OC) with resonant frequency
$\omega_c$. Such a system might be possible using the lumped-element circuit superconducting circuits with free standing drum-head capacitors recently developed by Teufel et al.~\cite{Teufel}. In fact, by adding an optical coating, the drum-head capacitor could also play the role of the micromirror of a Fabry-Perot optical cavity formed with a second standard input mirror. The microwave and optical cavities are driven at the
frequencies $\omega_{0w}=\omega_w-\Delta_{0w}$ and
$\omega_{0c}=\omega_c-\Delta_{0c}$, respectively. The Hamiltonian of
the coupled system reads~\cite{Vitali07,genes}
\begin{eqnarray}\label{ham0}
H&=&\frac{p_x^2}{2m}+\frac{m\omega^2_m
x^2}{2}+\frac{\Phi^2}{2L}+\frac{Q^2}{2[C+C_0(x)]}-e(t)Q\\
&&+\hbar\omega_c a^{\dagger}a-\hbar G_{0c}a^{\dagger}ax+i\hbar
E_c(a^{\dagger}e^{-i\omega_{0c}t}-a e^{i\omega_{0c}t}).\nonumber
\end{eqnarray}
where $(x,p_x)$ are the canonical position and momentum of a
MR with frequency $\omega_m$, $(\Phi,Q)$ are the
canonical coordinates for the MC, describing, the flux
through an equivalent inductor $L$ and the charge on an equivalent
capacitor $C$, respectively, $(a,a^{\dagger})$ show the annihilation
and creation operators of the OC
mode($[a,a^{\dagger}]=1$), $E_c=\sqrt{2P_c\kappa_c/\hbar\omega_{0c}}$
is related to input driving laser, where $P_c$ is the power of the
input laser and $\kappa_c$ describes the damping rate of the optical
cavity. $G_{0c}=(\omega_c/{\cal L})\sqrt{\hbar/m\omega_m}$ gives the optomechanical coupling rate, with $m$ the
effective mass of mechanical mode, and ${\cal L}$ the length
of the optical Fabry-Perot cavity, while the coherent
driving of the MC with damping rate $\kappa_w$ is given
by electric potential
$e(t)=-i\sqrt{2\hbar\omega_wL}E_w(e^{i\omega_{0w}t}-e^{-i\omega_{0w}t})$.
The MR is coupled to the microwave cavity because the capacity of the latter is a function of the resonator displacement,
$C_0(x)$. We expand this function around the equilibrium position of the resonator corresponding to a separation
$d$ between the plates of the capacitor, with corresponding bare capacitance $C_0$,
$C_0(x)=C_0[1+x(t)/d]$. Expanding the capacitive energy as a
Taylor series, we find to first order,
\begin{eqnarray}\label{2}
 \frac{Q^2}{2[C+C_0(x)]}=\frac{Q^2}{2C_{\Sigma}}-\frac{\mu}{2d
C_{\Sigma}}x(t)Q^2,
\end{eqnarray}
where $C_\Sigma=C+C_0$ and $\mu=C_0/C_\Sigma$.
The Hamiltonian of Eq.~(\eqn{ham0}) can be rewritten in the terms of the
raising and lowering operators of the MC field $b,
b^{\dagger}$($[b,b^{\dagger}]=1$) and the dimensionless position and
momentum operators of the MR, $\hat q$, $\hat p$ ($[\hat q,\hat
p]=i$), as
\begin{eqnarray}\label{ham1}
H&=&\hbar \omega_w b^{\dagger}b+\hbar \omega_c
a^{\dagger}a+\frac{\hbar \omega_m}{2}(\hat p^2+\hat q^2)-\frac{\hbar
G_{0w}}{2}\hat q(b+b^{\dagger})^2\nonumber\\
&&-\hbar G_{0c}\hat q a^{\dagger}a-i\hbar
E_w(e^{i\omega_{0w}t}-e^{-i\omega_{0w}t})(b+b^{\dagger})\nonumber\\
&&+i\hbar E_c(a^{\dagger}e^{-i\omega_{0c}t}-a e^{i\omega_{0c}t}),
\end{eqnarray}
where
\begin{eqnarray}\label{4}
b &=& \sqrt{\frac{\omega_w L}{2\hbar}}\hat Q+\frac{i}{\sqrt{2\hbar
\omega_w L}}\hat \Phi,\\
\hat q&=&\sqrt{\frac{m\omega_m}{\hbar}}\hat x,\;\;\;\;
\hat p=\frac{\hat p_{x}}{\sqrt{\hbar m\omega_m}},\\
G_{0w}&=&\frac{\mu \omega_w}{2d}\sqrt{\frac{\hbar}{m\omega_m}}.
\end{eqnarray}
It is then convenient to adopt the interaction picture with respect to
$H_0=\hbar \omega_{0w}b^{\dagger}b+\hbar \omega_{0c}a^{\dagger}a$, and neglect fast oscillating terms at $\pm2\omega_{0w}, \pm 2\omega_{0c}$, so that the system Hamiltonian becomes
\begin{eqnarray}\label{ham2}
H&=&\hbar \Delta_{0w} b^{\dagger}b+\hbar \Delta_{0c}
a^{\dagger}a+\frac{\hbar \omega_m}{2}(\hat p^2+\hat q^2)-\hbar
G_{0w}\hat qb^{\dagger}b\nonumber\\
&&-\hbar G_{0c}\hat q a^{\dagger}a-i\hbar E_w(b-b^{\dagger})+i\hbar
E_c(a^{\dagger}-a).
\end{eqnarray}
However the dynamics of the three modes is also affected by damping and noise processes, due to the fact that each of them interacts with its own environment. We can describe them adopting a QLE treatment in which the Heisenberg equations for the system operators associated with Eq.~(\ref{ham2}) are supplemented with damping and noise terms. The resulting nonlinear QLEs are given by
\begin{eqnarray}\label{lan1}
\dot{q}&=&\omega_m p,\\
\dot{p}&=&-\omega_m q-\gamma_m
p+G_{0c}a^{\dagger}a+G_{0w}b^{\dagger}b+\xi, \label{lan2}\\
\dot{a}&=&-(\kappa_c+i\Delta_{0c})a+iG_{0c}q
a+E_c+\sqrt{2\kappa_c}a_{in}, \label{lan3}\\
\dot{b}&=&-(\kappa_w+i\Delta_{0w})b+iG_{0w}q
b+E_w+\sqrt{2\kappa_w}b_{in}, \label{lan4}
\end{eqnarray}
where $\gamma_m$ is the mechanical damping rate and
$\xi(t)$ is the quantum Brownian noise acting on the MR, with
correlation function\cite{pin2,law}
\begin{equation}\label{nois1}
\langle\xi(t)\xi(t')\rangle=\frac{\gamma_m}{\omega_m}\int
\frac{d\omega}{2\pi}
e^{-i\omega(t-t')}\omega\left[\coth\left(\frac{\hbar \omega}{2k_B
T}\right)+1\right],
\end{equation}
where $k_B$ is the Boltzmann constant and $T$ is the temperature
of the reservoir of the mechanical resonator. We have also introduced the
optical and microwave input noises, respectively given by $a_{in}(t)$ and $b_{in}(t)$, obeying the following correlation functions \cite{gard}
\begin{eqnarray}\label{coropt}
\langle
a_{in}(t)a^{\dagger}_{in}(t')\rangle&=&[N(\omega_c)+1]\delta(t-t'),\\
\langle a^{\dagger}_{in}(t)a_{in}(t')\rangle&=&N(\omega_c)\delta(t-t'),\\
\label{cormicro}
\langle b_{in}(t)b^{\dagger}_{in}(t')\rangle&=&[N(\omega_w)+1]\delta(t-t'),\\
\langle b^{\dagger}_{in}(t)b_{in}(t')\rangle&=&N(\omega_w)\delta(t-t'),
\end{eqnarray}
where $N(\omega_c)=[\mathrm{exp}(\hbar \omega_c/k_B T)-1]^{-1}$ and
$N(\omega_w)=[\mathrm{exp}(\hbar \omega_w/k_B T)-1]^{-1}$ are the equilibrium
mean thermal photon number of the optical and microwave field,
respectively. One can safely assume $N(\omega_c)\approx0$ since $\hbar \omega_c/k_B T \gg 1$ at optical frequencies, while thermal microwave photons cannot be neglected in general, even at very low temperatures.

\section{Linearization of QLEs}

Our aim is to study the conditions under which one can efficiently correlate and entangle optical and microwave fields by means of the common interaction with a mechanical resonator. A straightforward way for achieving stationary and robust entanglement in continuous variable (CV) optomechanical systems, is to choose an operating point where the cavity is intensely driven so that the intracavity field is strong \cite{amo}. Under these conditions it is appropriate to focus onto the linearized dynamics of the quantum fluctuations around the
semiclassical fixed points. For this purpose, one can write
$a=\alpha_s+\delta a$, $b=\beta_s+\delta b$, $p=p_s+\delta p$, and
$q=q_s+\delta q$ and insert them into
Eqs.~(\eqn{lan1})-(\eqn{lan1}). The fixed points are obtained by setting the derivatives to zero, getting
\begin{eqnarray}\label{fixed1}
p_s &=& 0, \\
q_s &=& \frac{G_{0c}|\alpha_s|^2+G_{0w}|\beta_s|^2}{\omega_m},\label{fixed2}\\
\alpha_s & =& \frac{E_c}{\kappa_c+i\Delta_c},\label{fixed3}\\
\beta_s & =& \frac{E_w}{\kappa_w+i\Delta_w},\label{fixed4}
\end{eqnarray}
where $\Delta_c=\Delta_{0c}-G_{0c}q_s$ and
$\Delta_w=\Delta_{0w}-G_{0w}q_s$ describe the effective
detuning of the optical and microwave cavities field, respectively.
When both the optical and microwave intracavity fields are intense, $|\alpha_s|\gg 1$ and $|\beta_s|\gg 1$, one can safely linearize the dynamics around the steady state and obtain the following linear QLE for the quantum fluctuations of the tripartite system
\begin{subequations}\label{qles2}
\begin{eqnarray}
\delta \dot{q}&=&\omega_m \delta p,\\
\delta \dot{p}&=&-\omega_m \delta q-\gamma_m \delta p+G_{0c}\alpha_s (\delta
a^{\dagger}+\delta a)\nonumber \\
&&+G_{0w}\beta_s (\delta b^{\dagger}+\delta b)+\xi,\\
\delta \dot{a}&=&-(\kappa_c+i\Delta_c)\delta a+i G_{0c}\alpha_s \delta
q+\sqrt{2\kappa_c}a_{in},\\
\delta \dot{b}&=&-(\kappa_w+i\Delta_w)\delta b+i G_{0w}\beta_s \delta
q+\sqrt{2\kappa_w}b_{in},
\end{eqnarray}
\end{subequations}
where we have chosen the phase references so that $\alpha_s$ and
$\beta_s$ can be taken real and positive.

\section{Correlation matrix of the quantum fluctuations of the system}
The interaction of the optical and microwave field with the mechanical resonator is able to generate CV entanglement, i.e., quantum correlations between appropriate quadratures of the two intracavity fields and the position and momentum of the resonator. For this reason it is convenient to rewrite
Eqs.~(\eqn{qles2}) in terms of the OC field fluctuation
quadratures $\delta X_c=(\delta a+\delta a^{\dagger})/\sqrt{2}$ and
$\delta Y_c=(\delta a-\delta a^{\dagger})/i\sqrt{2}$, the microwave
cavity field fluctuation quadratures $\delta X_w=(\delta b+\delta
b^{\dagger})/\sqrt{2}$ and $\delta Y_w=(\delta b-\delta
b^{\dagger})/i\sqrt{2}$, and corresponding Hermitian input noise
operators $X^{in}_{c}=(\delta a_{in}+\delta
a_{in}^{\dagger})/\sqrt{2}$, $\delta Y^{in}_{c}=(\delta a_{in}-\delta
a_{in}^{\dagger})/i\sqrt{2}$, $X^{in}_{w}=(\delta b_{in}+\delta
b_{in}^{\dagger})/\sqrt{2}$, $\delta Y^{in}_{w}=(\delta b_{in}-\delta
b_{in}^{\dagger})/i\sqrt{2}$. The linearized QLE become
\begin{subequations}\label{qles3}
\begin{eqnarray}
\delta \dot{q}&=&\omega_m \delta p,\\
\delta \dot{p}&=&-\omega_m \delta q-\gamma_m p+G_{c}\delta
X_c+G_{w}\delta X_w+\xi,\\
\delta \dot{X_c}&=&-\kappa_c \delta X_c+\Delta_c \delta
Y_c+\sqrt{2\kappa_c}X^{in}_c,\\
\delta \dot{Y_c}&=&-\kappa_c \delta Y_c-\Delta_c \delta X_c+G_c \delta
q+\sqrt{2\kappa_c}Y^{in}_c,\\
\delta \dot{X_w}&=&-\kappa_w \delta X_w+\Delta_w \delta
Y_w+\sqrt{2\kappa_w}X^{in}_w,\\
\delta \dot{Y_w}&=&-\kappa_w \delta Y_w-\Delta_w\delta X_w+G_w \delta
q+\sqrt{2\kappa_w}Y^{in}_w,
\end{eqnarray}
\end{subequations}
where
\begin{eqnarray}\label{g1}
G_c&=&\sqrt{2}G_{0c}\alpha_s=\frac{2\omega_c}{L}\sqrt{\frac{P_c\kappa_c}{m
\omega_m \omega_{0c}(\kappa_c^2+\Delta_c^2)}},\\
\label{g2}
G_w&=&\sqrt{2}G_{0w}\beta_s=\frac{\mu
\omega_w}{d}\sqrt{\frac{P_w\kappa_w}{m \omega_m
\omega_{0w}(\kappa_w^2+\Delta_w^2)}},
\end{eqnarray}
are the effective coupling of the optical and microwave cavity fluctuations with the MR.
The Eqs.(\eqn{qles3}) can be written in the following matrix
form
\begin{equation}\label{drift}
\begin{array}{rcl}
\dot u(t)=A u(t)+n(t),
\end{array}
\end{equation}
where $u^T(t)=[\delta q(t),\delta p(t),\delta X_c(t),\delta
Y_c(t),\delta X_w(t),\delta Y_w(t)]\,\,$\\
(the superscript $T$ denotes the transposition),
$n^T(t)=[0,\xi(t),\sqrt{2\kappa_c} X^{in}_c,\sqrt{2\kappa_c}
Y^{in}_c,\sqrt{2\kappa_w} X^{in}_w,\sqrt{2\kappa_w} Y^{in}_w]$, and
\begin{equation}\label{driftA}
\begin{array}{rcl}
A = \left( {\begin{array}{*{20}c}
   0 & {\omega _m } & 0 & 0 & 0 & 0  \\
   { - \omega _m } & { - \gamma _m } & {G_c } & 0 & {G_w } & 0  \\
   0 & 0 & { - \kappa _c } & {\Delta _c } & 0 & 0  \\
   {G_c } & 0 & { - \Delta_c } & { - \kappa _c } & 0 & 0  \\
   0 & 0 & 0 & 0 & { - \kappa _w } & {\Delta _w }  \\
   {G_w } & 0 & 0 & 0 & { - \Delta _w } & { - \kappa _w }  \\
\end{array}} \right).
\end{array}
\end{equation}
If all the eigenvalues of the drift matrix $A$ possess negative real parts, the system is stable and reaches a steady state. The stability conditions can be explicitly derived e.g., by means of the Routh-Hurwitz criteria \cite{routh}, whose explicit expression is however quite cumbersome and will not be reported here. Due to the Gaussian nature of the quantum noise terms in Eq.~(\eqn{drift}) and the linear dynamics,
the steady state of the quantum fluctuations of the
system is a CV tripartite Gaussian state, completely
characterized by the $6\times 6$ correlation matrix (CM) with
corresponding components $V_{ij}=\langle
u_i(\infty)u_j(\infty)+u_j(\infty)u_i(\infty)\rangle/2 $. When the system is stable such a CM is given by
\begin{equation}\label{vmatrix}
V_{ij}=\sum_{k,l}\int_{0}^{\infty}ds\int_{0}^{\infty}ds'M_{ik}(s)M_{jl}(s')\Phi_{kl}(s-s'),
\end{equation}
where $M(s)=\exp(As)$ and $\Phi_{kl}(s-s')=\langle
n_k(s)n_l(s')+n_l(s')n_k(s)\rangle /2 $ is the matrix of stationary
noise correlation functions.
The noise term, $\xi(t)$,  is not $\delta$-correlated and therefore
does not describe a Markovian process. Typically, significant optomechanical entanglement is
achieved for a very high mechanical quality factor,
$Q=\omega_m/\gamma_m\rightarrow \infty$. In this limit, $\xi(t)$
becomes with a good approximation delta-correlated \cite{kac}, i.e.,
$\langle\xi(t)\xi(t')+\xi(t')\xi(t)\rangle \simeq 2\gamma_m(2\bar{n}+1)\delta
(t-t')$, where $\bar{n}=[\mathrm{exp}(\hbar \omega_m/k_B T)-1]^{-1}$ is the
mean thermal excitation number of the resonator. As a consequence,
$\Phi_{kl}(s-s')=D_{kl}\delta(s-s')$, where
$D_{kl}=\mathrm{Diag}[0, \gamma_m(2\bar{n}_b+1),\kappa_c,\kappa_c,\kappa_w(2N(\omega_w)+1),\kappa_w(2N(\omega_w)+1)],
$ is the diffusion matrix, so that Eq.~(\eqn{vmatrix}) becomes
 \begin{equation}\label{vmatrix2}
V=\int_{0}^{\infty}dsM(s)D M^T(s).
\end{equation}
When the system is stable $M(\infty)=0$ and Eq.~(\eqn{vmatrix2})
is equivalent to the following Lyapunov equation for the steady-state CM
\begin{equation}\label{lyap}
AV+VA^T=-D,
\end{equation}
which is linear in $V$ and can be straightforwardly solved. However, its explicit solution is cumbersome and
will not be reported here.

We are interested in the entanglement properties of the steady state of the tripartite system under study and therefore we shall focus onto the entanglement of the three possible bipartite subsystems that can be formed by tracing over the remaining degree of freedom. Such bipartite entanglement will be quantified using the logarithmic negativity \cite{eis},
 \begin{equation}\label{en}
E_N=\mathrm{max}[0,-\mathrm{ln} 2 \eta^-],
\end{equation}
where $\eta^{-}\equiv2^{-1/2}\left[\Sigma(V_{bp})-\sqrt{\Sigma(V_{bp})^2-4 \mathrm{det} V_{bp}}\right]^{1/2}$  is the lowest symplectic eigenvalue of the partial transpose of the $4 \times 4$ CM, $V_{bp}$, associated with the selected bipartition, obtained by neglecting the rows and columns of the uninteresting mode,
\begin{equation}\label{loga}
V_{bp}=\left(
     \begin{array}{cc}
       B & C \\
       C^T & B' \\
     \end{array}
   \right),
\end{equation}
and $\Sigma(V_{bp})\equiv \mathrm{det} B+\mathrm{det} B'-2\mathrm{det} C$.

\section{Results}

The results are shown in Figs.~2 and 3, where we study the dependence of the three possible bipartite entanglement at the steady state of the system versus the detuning of the microwave cavity and for different values of the temperature and of the mass of the mechanical resonator. As suggested in the introduction, we have assumed an experimental situation representing a feasible extension of the scheme of Ref.~\cite{Teufel}, i.e., we have assumed a lumped-element superconducting circuit with a free standing drum-head capacitor which is then optically coated to form a micromirror of an additional optical Fabry-Perot cavity. We have taken parameter analogous to that of Ref.~\cite{Teufel} for the MC and MR, that is,
a MR with $\omega_m/2\pi=10$ MHz, $Q=5\times 10^4$, and a MC with $\omega_w/2\pi=10$ GHz, $\kappa_w=0.02\omega_m$, driven by a microwave source with power
$P_w=30$ mW. The coupling between the two is determined by the parameters $d=100$ nm, $\mu=0.008$. We have considered a lower mechanical quality factor, and resonator masses between $m = 10$ ng and $m = 100$ ng, i.e., higher than that of Ref.~\cite{Teufel}, in order to take into account the presence of the coating, which typically worsens the mechanical properties. We have then assumed an OC of length $L=1$ mm, damping
rate $\kappa_c=0.08\omega_m$, driven by a laser with wavelength $\lambda_{0c}=810$ nm and power $P_c=30mW$. The optical cavity detuning has been fixed at $\Delta_c = \omega_m$ which turns out to be the most convenient choice (see below).

\begin{figure}[ht]
\includegraphics[width=0.55\textwidth]{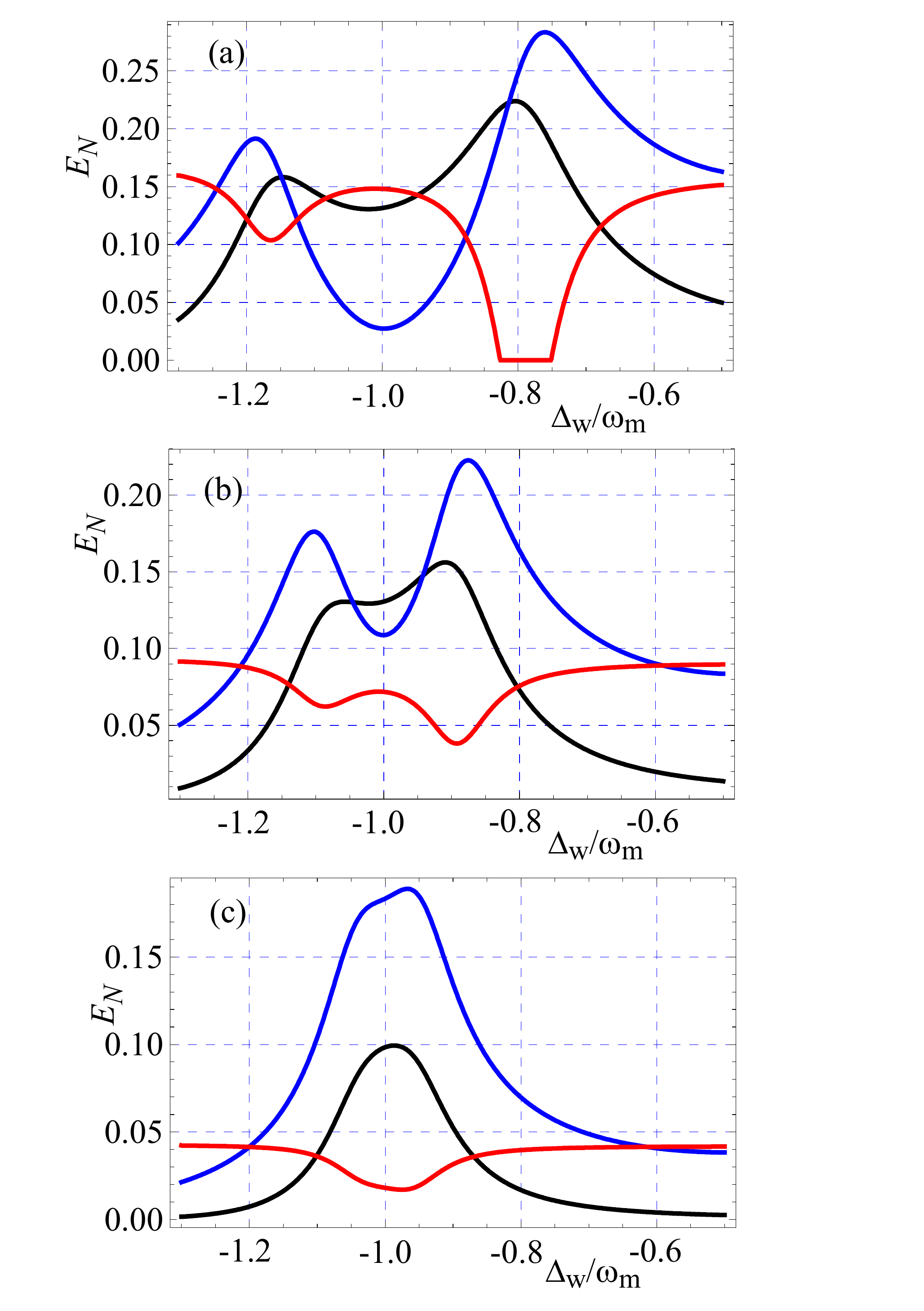}
\caption{(Color online) Plot of $E_{N}$ of the three bipartite subsystems (OC-MC full black line, OC-MR dotted red line, MC-MR dashed blue line) versus the normalized microwave cavity detuning $\Delta_w/\omega_{m}$ at fixed temperature $T=15$ mK, and at three different MR masses: $m=10$ ng (a), $m=30$ ng (b), $m=100$ ng (c). The optical cavity detuning has been fixed at $\Delta_c = \omega_m$, while the other parameters are $\omega_m/2\pi=10$ MHz, $Q=5\times 10^4$, $\omega_w/2\pi=10$ GHz, $\kappa_w=0.02\omega_m$, $P_w=30$ mW, $d=100$ nm, $\mu=0.008$, $\kappa_c=0.08\omega_m$, $P_c=30mW$.}
\label{fig:2}
\end{figure}

\begin{figure}[ht]
\includegraphics[width=0.55\textwidth]{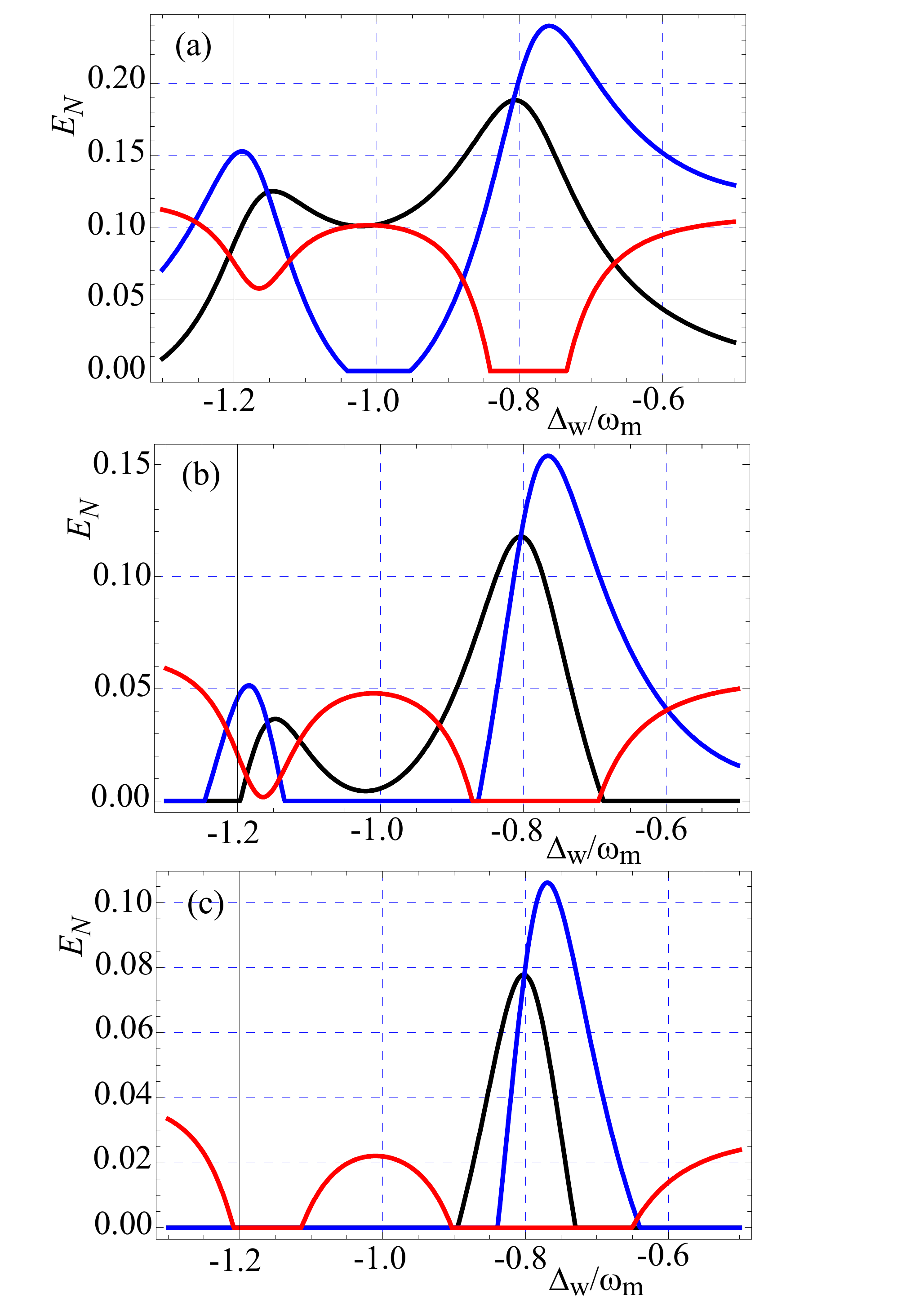}
\caption{(Color online) Plot of $E_{N}$ of the three bipartite subsystems (OC-MC full black line, OC-MR dotted red line, MC-MR dashed blue line) versus the normalized microwave cavity detuning $\Delta_w/\omega_{m}$ at fixed resonator mass $m=10$ ng and at three different temperatures: $T=100$ mK (a), $T=200$ mK (b), $T=250$ mK (c). The optical cavity detuning has been fixed at $\Delta_c = \omega_m$, while the other parameters are as in Fig.~2.}
\label{fig:3}
\end{figure}

Fig.~2 shows the three bipartite logarithmic negativities, $E_N^{wc}$ (full black line), $E_N^{wm}$ (dashed blue line), $E_N^{mc}$ (dotted red line)
as a function of the normalized microwave cavity detuning $\Delta_w/\omega_{m}$ at fixed temperature $T=15$ mK, and at three different MR masses: $m=10$ ng (a), $m=30$ ng (b), $m=100$ ng (c). We see that for increasing masses the three kind of bipartite entanglement decrease overall, because by increasing masses, both couplings $G_c$ and $G_w$ decrease. However the three logarithmic negativities do not behave in the same way and entanglement sharing is evident. In particular the entanglement of interest, i.e., $E_N^{wc}$, increases at the expense of the optical-mechanical entanglement, while $E_N^{wm}$ remains always non-negligible. Another relevant feature (which we explain in the following) are the double peaks of $E_N^{wc}$ and $E_N^{wm}$, which however tends to merge for larger masses, i.e., smaller couplings.

The stability conditions are always satisfied in the chosen parameter regime, which corresponds to a situation where the red-detuned cavity field (the optical mode in the present case) is coupled more strongly to the resonator than the blue-detuned field (the microwave mode), because instability is achieved easier in the blue-detuned case \cite{amo}.

Fig.~3 instead shows the log-negativity of the three bipartite subsystems (OC-MC full black line, OC-MR dotted red line, MC-MR dashed blue line) versus the normalized microwave cavity detuning $\Delta_w/\omega_{m}$ at fixed resonator mass $m=10$ ng, and at three different temperatures: $T=100$ mK (a), $T=200$ mK (b), $T=250$ mK (c). As expected, the three kinds of bipartite entanglement decrease at increasing temperatures, and even if nonzero, the log-negativity is already quite small at $T=200$ mK. Also in these plots entanglement sharing is evident and one kind of entanglement always increases at the expense of the others. More in general, the scheme is able to generate appreciable entanglement between the optical and the microwave cavity fields, especially at the expense of the optical-mechanical entanglement, while unfortunately the mechanical resonator remains always appreciably entangled with the microwave cavity. Similar bipartite entanglement dynamics can be observed in other alike tripartite system, as
the atom-field-mirror system of Ref.~\cite{fam} and the two cavity optomechanical setup of Ref.~\cite{paternostro}.

One can provide an intuitive explanation of the above behavior and in particular of the mechanism through which the optical and microwave modes gets entangled through their interactions with the mechanical resonator. It is convenient to move to the interaction picture with respect to $H_{\Delta}=\hbar \Delta_{c} a^{\dagger} a+\hbar \Delta_{w} b^{\dagger} b$, formally solve the dynamics of the MR in Eqs.~(\ref{qles2}), and insert this formal solution into the dynamical equations of the two modes. One gets the following exact equations
\begin{subequations}\label{qlesappr}
\begin{eqnarray}
&&\delta \dot{a}=-\kappa_c \delta a+\sqrt{2\kappa_c}a_{in}(t)e^{i\Delta_c t}\\
&&+\frac{i}{2} \int_{-\infty}^t ds \chi_M(t-s)\left\{G_c \xi(s)e^{i\Delta_c t} \right. \nonumber \\
&&\left.+G_c^2\left[\delta a(s)e^{i\Delta_c (t-s)} +\delta a^{\dagger}(s)e^{i\Delta_c (t+s)}\right]\right. \nonumber \\
&&\left.+ G_c G_w\left[\delta b(s)e^{i\Delta_c t-i\Delta_w s}+ \delta b^{\dagger}(s)e^{i\Delta_c t+i\Delta_w s}\right]\right\},\nonumber \\
&&\delta \dot{b}=-\kappa_w \delta b+\sqrt{2\kappa_w}b_{in}e^{i\Delta_w t} \\
&&+\frac{i}{2} \int_{-\infty}^t ds \chi_M(t-s)\left\{G_w \xi(s) e^{i\Delta_w t}\right. \nonumber \\
&&\left.+G_w^2\left[\delta b(s)e^{i\Delta_w (t-s)} +\delta b^{\dagger}(s)e^{i\Delta_w (t+s)}\right]\right.\nonumber \\
&&\left.+ G_c G_w\left[\delta a(s)e^{i\Delta_w t-i\Delta_c s}+ \delta a^{\dagger}(s)e^{i\Delta_w t+i\Delta_c s}\right]\right\}, \nonumber
\end{eqnarray}
\end{subequations}
where
\begin{equation}\label{eq:susc}
   \chi_M(t)=\int_{-\infty}^{\infty} \frac{d\omega}{2\pi}\frac{e^{-i \omega t}\omega_m}{ \omega_m^2-\omega^2-i \gamma_m \omega}\simeq e^{-\gamma_m t/2} \sin \omega_m t
\end{equation}
is the mechanical susceptibility. These equations show that the MR mediates an effective retarded interaction between the optical and cavity modes which may originate various phenomena: i) cavity frequency shifts and single mode squeezing for both modes; ii) excitation transfer between the two modes; iii) two-mode squeezing between optical and microwave photons. One can resonantly select one of these processes by appropriately adjusting the cavity detunings. In particular, if we choose opposite detunings $\Delta_c = -\Delta_w \equiv \Delta \simeq \omega_m$, and assume the regime of fast mechanical oscillations, $\Delta \sim \omega_m \gg G_c,G_w,\kappa_c,\kappa_w$, which allows us to neglect the fast oscillating terms at $\sim \pm 2 \Delta $ in the above equations, single mode squeezing and excitation transfer terms become negligible and Eqs.~(\ref{qlesappr}) can be written as
\begin{subequations}\label{qlesappr2}
\begin{eqnarray}
&&\delta \dot{a}=-\kappa_c \delta a+\sqrt{2\kappa_c}\tilde{a}_{in}(t)+\frac{i}{2} \int_{-\infty}^t ds \chi_M^{\Delta}(t-s)\left\{G_c \tilde{\xi}(s)\right. \nonumber\\
&&\left.+G_c^2 \delta a(s)+ G_c G_w \delta b^{\dagger}(s)\right\}, \\
&&\delta \dot{b}=-\kappa_w \delta b+\sqrt{2\kappa_w}\tilde{b}_{in}(t)+\frac{i}{2} \int_{-\infty}^t ds \chi_M^{\Delta}(t-s)^*\left\{G_w \tilde{\xi}^{\dagger}(s) \right. \nonumber \\
&&\left.+G_w^2 \delta b(s)+ G_c G_w \delta a^{\dagger}(s)\right\},\nonumber \\
\end{eqnarray}
\end{subequations}
where we have redefined the noise terms $\tilde{\xi}(t)=\xi(t)e^{i\Delta t}$, $\tilde{a}_{in}(t)=a_{in}(t)e^{i\Delta_c t}$, $\tilde{b}_{in}(t)=b_{in}(t)e^{i\Delta_w t}$, and also $\chi_M^{\Delta}(t)=\chi_M(t)e^{i\Delta t}$. These equations show
that apart from noise terms and frequency shifts, the two modes undergo a retarded parametric interaction with a time dependent coupling kernel $G_c G_w\chi_M^{\Delta}(t)/2$. This parametric interaction is resonantly large when $\Delta \sim \omega_m$ because when this condition is not satisfied the kernel rapidly oscillates and the interaction tends to average to zero. Therefore for not too large $G_c$ and $G_w$, one expects to find optical-microwave entanglement around $\Delta_w=-\Delta_c=\omega_m$, as it is evident in Fig.~2c corresponding to a larger mass and therefore smaller couplings. For smaller masses, the rotating-wave argument above starts to be less valid, but one expects that the ``resonance'' condition for entanglement $\Delta_w = -\Delta_c$ is replaced by a condition $\Delta_w = -\lambda_{\pm}$, where $\lambda_{\pm}$ are the two eigenvalues corresponding to normal-mode splitting \cite{nms}, i.e., to the splitting of the two degenerate eigenvalues which are equal to $\Delta_c=\omega_m$ in the limit of very small couplings. This normal-mode splitting phenomenon explains therefore the double-peak structure of the optical-microwave entanglement appearing at smaller masses.

\section{CONCLUSIONS}
We have proposed a scheme for the realization
of a hybrid quantum correlated tripartite system formed
by an optical cavity and a microwave cavity, both interacting with a mechanical resonator. We have studied its dynamics adopting a QLE treatment. We have focused on the steady state of the system and in particular on the stationary quantum fluctuations of the system, by solving the linearized dynamics around the classical steady state. We have seen that in an experimentally accessible parameter
regime, the steady state of the system shows bipartite CV entanglement for all the possible bipartite subsystems. In particular the MR is able to mediate the realization of robust stationary (i.e., persistent) entanglement between the optical and microwave cavity modes, which could be extremely useful in quantum information network in which light modes are used for quantum communications, and microwave cavities are used in the nodes for interfacing with solid-state qubits. In fact, in quantum networks one often needs to transfer and store states carried by light modes into local quantum processors, which are typically manipulated by microwave fields. The hybrid device discussed here, thanks to the mediating action of the MR, allows to establish a robust and stationary entanglement between the optical and microwave fields which can be exploited for such a state transfer. For example, high-fidelity state transfer from an optical field onto the intracavity microwave field can be realized by implementing a CV teleportation protocol in which the input optical field and the optical cavity field are subject to homodyne measurements and the result of this measurement is fed-forward to the microwave cavity, where it is used for conditionally displacing its state. The present values of $E_N$, even if not large, allows to achieve teleportation fidelities larger than those achievable with only classical means (see for example \cite{ade-illu05,Mari2008}).

\section{Acknowledgments}

This work has been supported by the European Commission (FP-7 FET-Open project MINOS and HIP).


\end{document}